\documentclass[12pt, draftclsnofoot, onecolumn]{IEEEtran}
\def\BibTeX{{\rm B\kern-.05em{\sc i\kern-.025em b}\kern-.08em T\kern-.1667em\lower.7ex\hbox{E}\kern-.125emX}}
\usepackage{graphicx}
\usepackage{amsmath,bbm,epsfig,amssymb,amsfonts, amstext, verbatim,amsopn,cite,subfigure,multirow,multicol,lipsum}
\usepackage{balance}
\usepackage{url}
\usepackage{amsfonts}
\usepackage{epsfig}
\usepackage{epstopdf}
\usepackage{setspace}
\usepackage{stmaryrd}
\usepackage{psfrag}	
\usepackage{multirow}
\usepackage{float}
\usepackage{cases}  
\usepackage[process=auto]{pstool}
\usepackage{etoolbox}
\usepackage{hyperref}
\usepackage{wasysym}[nointegrals]
\usepackage{pifont}
\allowdisplaybreaks
\usepackage{graphicx}
\usepackage{wrapfig}
\usepackage{lscape}
\usepackage{rotating}
\usepackage{epstopdf}

\newtheorem{theo}{Theorem}
\newtheorem{lem}{Lemma}

\newtheorem{corol}{Corollary}

\newtoggle{Conf}

\togglefalse{Conf}

\iftoggle{Conf}{%
}{%
}

\EndPreamble
\begin{document}

\title{  Channel Modeling for IRS-Assisted FSO Systems
\vspace{-0.3cm}}
\author{Hedieh Ajam, Marzieh Najafi, Vahid Jamali,  and Robert Schober\\  
Friedrich-Alexander University  Erlangen-Nuremberg, Germany
\vspace{-0.2cm}}
\maketitle
\begin{abstract}
In this paper, we develop an analytical channel model for intelligent reflecting surface (IRS)-assisted  free space optical (FSO) systems.  Unlike IRS-assisted radio frequency systems, where it is typically assumed that a plane wave is incident  on the IRS, in FSO systems, the incident wave is a Gaussian beam with non-uniform power distribution  across the IRS. Taking this property  into account, we develop an analytical end-to-end channel    model for IRS-assisted FSO systems  based on the Huygens-Fresnel principle. Our analytical  model reveals the impact of the size, position,  orientation, and phase-shift configuration of the IRS  on the end-to-end channel.   Furthermore, we show that  results obtained based on geometric optics under the far-field approximation are only valid for a specific   range of  IRS-receiver lens distances depending on the IRS size, incident beam width, and wavelength.    Simulation results validate the accuracy of the proposed analytical results for the   FSO beam reflected from the IRS and compare the bit error rate performance obtained for the proposed analytical channel model with that obtained for  geometric optics under the far-field approximation.
\end{abstract}
\section{Introduction}
Metamaterials can manipulate the  properties of a wave such as its polarization, phase, and amplitude in reflection  and  transmission \cite{Arbabi}, \cite{Marco_survey}.  Intelligent reflecting surfaces (IRSs) are  planar structures of metamaterials with subwavelength thickness and consist of many subwavelength elements referred to as unit cells \cite{Arbabi}.  In order to achieve   a desired behavior, the IRS  changes the accumulated phase of the   wave   reflected by the surface.  The phase shifts applied by the  unit cells  determine the characteristics of the IRS. In the literature, typically    a linear phase shift gradient is assumed, so that the IRS can provide  anomalous reflection of a beam in a desired direction \cite{Vahid_IRS,Estakhri,Kochkina}. In radio frequency (RF) wireless  communication systems, IRS have been exploited to increase coverage, ensure security, harness interference, and  improve the quality of  non-line-of-sight (NLOS) connections \cite{zhang}. For free   space  optical (FSO) systems, which require in general a line-of-sight (LOS) connection,  the authors of \cite{Marzieh_IRS}  employed an optical IRS to connect  a transmitter  with an obstructed receiver.  Unlike  RF waves,  which have a  plane wavefront  and uniform power distribution across the IRS, the    Gaussian beams employed in FSO systems have a curved wavefront and a non-uniform power distribution.  The authors of \cite{Marzieh_IRS} exploited geometric optics using a far-field approximation to determine the impact of  IRSs on the performance of  FSO systems. In this paper, we will show that this approach is only valid for  specific IRS-receiver lens distances, incident beam widths, and IRS sizes.   Moreover, the authors of  \cite{practicalIRS} applied an optical IRS to enhance an indoor communication link. In \cite{Alouini_IRS},  the impact of  IRSs on   visible light communications (VLC) was investigated. However,   VLC employs  non-directional beams which exhibit a  different behavior compared to the   Gaussian laser beams used in  FSO systems.

In this paper, we employ  an  IRS-assisted  FSO system to provide an LOS connection between a transmitter and an obstructed  receiver. The transmitter is equipped with a laser source (LS) emitting a Gaussian  beam which is reflected by an IRS towards the receiver where the beam is focused by a lens onto a photo detector (PD). Based on  the Huygens-Fresnel principle, we analyze the channel gain of this IRS-assisted FSO system  taking the non-uniform power distribution  of  the Gaussian beam into account.    In the following, we summarize the main contributions of our work.
\begin{itemize}
	\item  Based on the Huygens-Fresnel principle, we derive the deterministic channel gain  of  an  IRS-assisted FSO system   which employs a Gaussian beam  that is emitted by an LS and is reflected by   an optical IRS. The  impact of   the relative position of the IRS with respect to (w.r.t.)  the lens and the LS,   the size of the IRS, and the phase-shift configuration of the IRS are included in the proposed analytical end-to-end channel model.   
	\item  In  the  far-field approximation, the wave reflected  by the IRS is modeled as parallel rays with linear phase differences. However, we will show that depending on the IRS size and the incident beam width, this approximation  may not always be valid.  We mathematically characterize  the  range of intermediate and far-field distances and	propose an analytical channel  model that is valid for these distances.  We  show that large distances (e.g. more than $32$ km for a square-shaped  IRS of length $0.5$ m and incident beam widths of $0.19$ m and $0.52$ m) may be needed for the far-field approximation to be valid.
	\item Our simulation results validate the analytical results for the electric field of the beam reflected by the IRS and the deterministic  gain of the IRS-assisted FSO channel. Moreover, we compare the bit error rate (BER) performance  of an IRS-assisted FSO system for the developed analytical channel  model with   results obtained for the channel model based on geometric optics under the far-field approximation. Our simulation results confirm that for larger IRS sizes and beam widths and practical distances, the far-field approximation is not valid and the proposed model has to  be used.   
\end{itemize}

\textit{Notations:} Boldface lower-case and upper-case letters denote vectors and matrices, respectively.  Superscript $(\cdot)^T$ and $\mathbb{E}\{\cdot\}$
denote the transpose and expectation operators, respectively. $\mathbf{I}_n$ is the $n\times n$ identity matrix, $j$ denotes the imaginary unit, and $\text{conj}(\cdot)$ and  $\text{real}(\cdot)$ represent the complex conjugate and the real part of a complex number, respectively. Moreover, $\text{erf}(\cdot)$ and $\text{erfi}(\cdot)$ are the error function and the imaginary error function, respectively.  
\section{System Model}\label{Sec_System}
\begin{figure}
	\centering
	\includegraphics[width=0.55\textwidth]{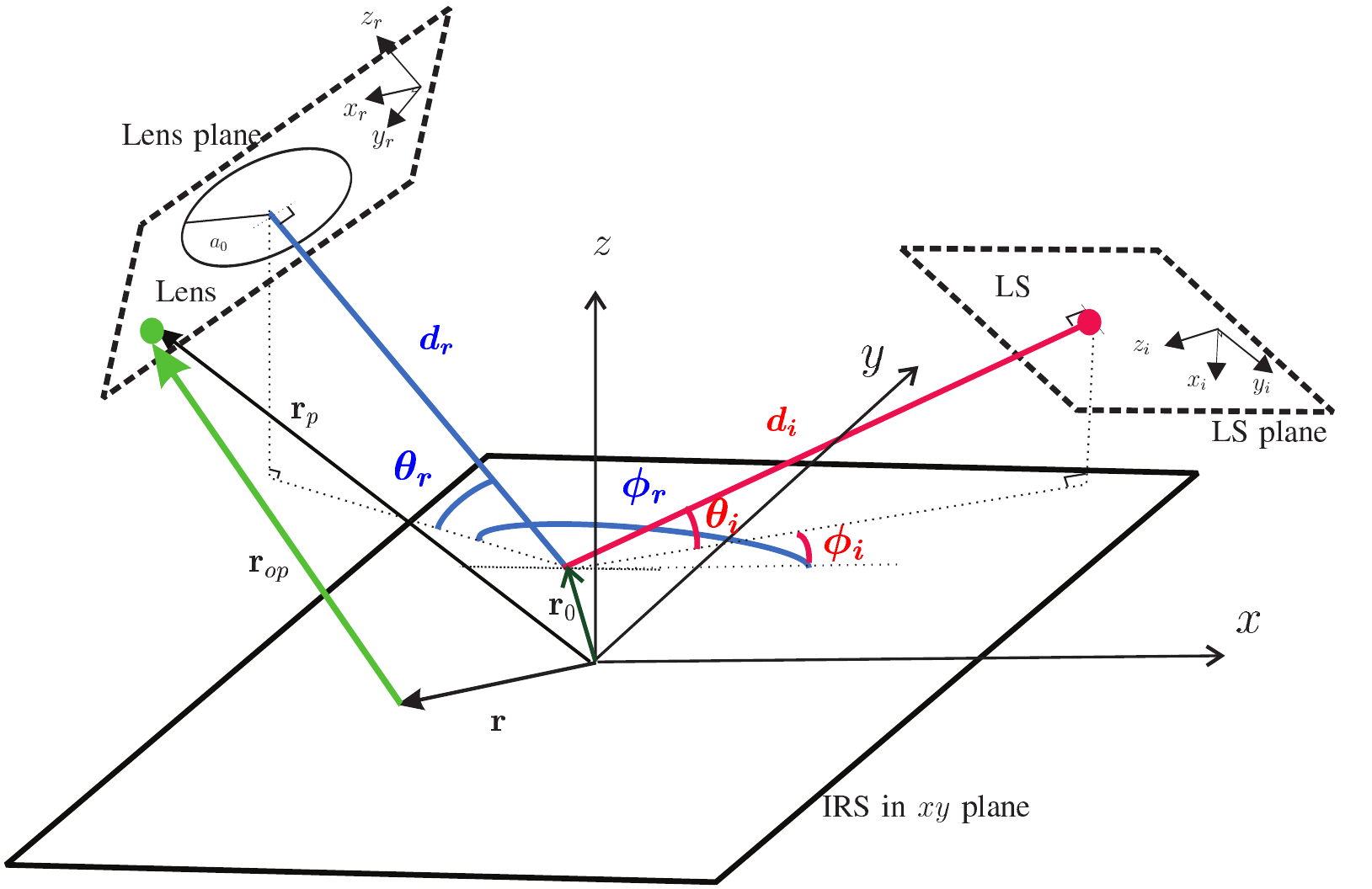}
	\caption{Schematic model of IRS-assisted FSO system.}
	\label{Fig:System}\vspace{-0.3cm}
\end{figure}

We consider an  IRS-assisted FSO system where a transmitter is equipped with an LS to communicate through an IRS with a  receiver equipped with a PD and a lens, see Fig.~\ref{Fig:System}. In particular, the LS illuminates the IRS with a Gaussian laser beam and the receiver collects the  power reflected by the IRS with its lens and detects the received power with its PD \cite{Marzieh_IRS}.  The IRS is positioned in the $xy$-plane and centered at the origin. It has length   $L_x$ in  $x$-direction and width $L_y$ in  $y$-direction. The IRS consists of many subwavelength unit cells and by properly adjusting their phases, the IRS is able to reflect the incident beam from the LS towards the  receiver. Given that  the IRS is larger than the beam wavelength, i.e., $L_x, L_y \gg \lambda$, it can be modeled as a continuous surface with continuous phase-shift configuration, denoted by $\Phi_\text{irs}$ \cite{Vahid_IRS}. Here, we assume  a linear phase gradient across the unit cells, similar to previous related works  \cite{Marzieh_IRS_jou,Vahid_IRS}, see Section \ref{Sec_Analysis} for more details. In the considered system, the  LS is located at $\mathbf{p}_{i}=(d_i, \theta_i, \phi_i)$, where $d_i$ is the distance between  the LS and the beam footprint center on the IRS along the beam axis, $\theta_i$ is the angle between the $xy$-plane and the beam axis, and $\phi_i$ is the angle between the projection of the beam axis on the $xy$-plane and the  $x$-axis.   Without loss of generality, we assume $\phi_i=0$ in the following. Moreover,   the beam footprint on the IRS plane is centered at point $\mathbf{r}_0=\left(x_{0}, y_{0}, 0\right)$. Furthermore, we assume the receiver is equipped with a circular lens of radius $a_0$ which is always perpendicular to the received  beam axis and  the lens center is positioned at $\mathbf{p}_{r}=(d_r, \theta_r, \phi_r)$, where $d_r$ is the distance between the beam footprint center on the  IRS and the lens center, $\theta_r$ is the angle between the $xy$-plane and the normal vector of the lens plane, and $\phi_r$ is the angle between the projection of the normal vector of the lens plane on the $xy$-plane and the $x$-axis. 

At the receiver, the lens focuses the  beam, which was reflected by the IRS, onto the PD and the  signal  $y$ received at the PD can be modeled as
\begin{IEEEeqnarray}{rll}
	y=h  s +n, 
\end{IEEEeqnarray}
where  $s$ is an on-off keying (OOK) modulated optical symbol  with average power $\sigma_{s}^2=\mathbb{E}\{|s|^2\}$, $h\in \mathcal{R}^+$ is the channel gain, and $n$ is additive white Gaussian noise (AWGN) with zero mean and variance $\sigma_n^2$.
In general, FSO channels are affected by  geometric and misalignment losses,  atmospheric losses, and  atmospheric turbulence \cite{ICC}. For simplicity, we assume  perfect beam tracking and  that the positions of the IRS,  LS, and  lens do not fluctuate. Thus, the misalignment loss can be ignored, see \cite{Marzieh_IRS} for an analysis of the misalignment loss of  IRS-assisted FSO systems. Then, the IRS-assisted FSO channel gain $h$ between the  LS  and the PD  can be modeled as
\begin{IEEEeqnarray}{rll}
	h=h_p h_{\text{irs}} h_a,
	\label{ch}
\end{IEEEeqnarray}
where $h_a$ is the random atmospheric turbulence component,  $h_p=10^{-\frac{\kappa}{10}(d_i+d_r)}$ is the atmospheric loss dependent on attenuation factor, $\kappa$, and $h_{\text{irs}}$ is  the geometric  loss. We assume that $h_a$ is Gamma-Gamma distributed, i.e., $h_a\sim \mathcal{GG}(\alpha, \beta)$, where $\alpha$ and $\beta$ are the small and large scale turbulence parameters \cite{GG_Murat}. Moreover, $h_\text{irs}$ denotes the fraction of power  reflected by the IRS and collected by the PD and is given by
\begin{IEEEeqnarray}{rll}
	h_\text{irs}=\iint\limits_{(x_r, y_r)\in \mathcal{A}} I_\text{irs}\left(\mathbf{r}_r \right)\, \mathrm{d}x_r \mathrm{d}y_r,
	\label{gain}
\end{IEEEeqnarray}
where $\mathcal{A}$ is the area of the lens and $\mathbf{r}_r=(x_r, y_r, z_r)$ denotes the   coordinates   on the lens plane.  The origin of the $x_ry_rz_r$-coordinate system  is located at the center of the lens and the $z_r$-axis  points in the opposite direction of  the normal vector of the lens plane, see Fig.~\ref{Fig:System}. We assume that the $y_r$-axis is along the intersection line of the lens plane and the IRS plane and the $x_r$-axis is perpendicular to the $y_r$-axis.  Here, $I_\text{irs}(\mathbf{r}_r)$ is  the power density of the reflected beam  in  the  lens plane.  Given that the Gaussian laser beam  waist, $w_0$, is larger than the wavelength, $\lambda$, the paraxial approximation is valid and the power density of the  reflected beam can be expressed as follows  \cite{saleh}
\begin{IEEEeqnarray}{rll}
	I_\text{irs}(\mathbf{r}_r)=\frac{1}{2\eta}\lvert E_r(\mathbf{r}_r) \rvert^2,
	\label{irradiance}
\end{IEEEeqnarray}
where $\eta$ is the free-space impedance and $E_r(\mathbf{r}_r)$ is the  electric field reflected by the IRS and observed at the lens. 
 
The electric field of the Gaussian laser beam  emitted by the LS   is given by \cite{Goodmanbook}
\begin{IEEEeqnarray}{rll}
	E_o(\mathbf{r}_i)=& \frac{E_0w_0}{{w}(z_i)} \exp\left(-{x_i^2+y_i^2\over w^2(z_i)}-j\phi_o\right)\quad\text{with phase} \nonumber\\
				\phi_o&=k\left(z_i+{x_i^2+y_i^2\over 2R(z_i)}\right)-\tan^{-1}\left(\frac{z_i}{z_0}\right),\quad
	\label{Gauss}
\end{IEEEeqnarray}
where $\mathbf{r}_i=(x_i,y_i,z_i)$ is a point in a coordinate system which has its origin at the LS, the $z_i$-axis  is along the beam axis, the $y_i$-axis is along the intersection line of the LS plane and the IRS plane, and the $x_i$-axis  is orthogonal to the $y_i$-axis. Here, $E_0$ is the  electric field at the origin, $k=\frac{2\pi}{\lambda}$ is the wave number, $w(z_i)=w_0\left[{1+\left(\frac{z_i}{z_0}\right)^2}\right]^{1/2}$ is the beam width at distance $z_i$, $R(z_i)=z_i\left[1+\left(\frac{z_0}{z_i}\right)^2\right]$ is the radius of curvature of the beam's wavefront, and $z_0=\frac{\pi w_0^2}{\lambda}$ is the Rayleigh range.

\section{Proposed Channel Model}\label{Sec_Analysis}
Here, given the Gaussian  laser beam in (\ref{Gauss}), first  we  determine the   electric field incident on the IRS, then,   we determine the reflected electric field $E_{r}(\mathbf{r}_r)$ using the Huygens-Fresnel principle. Next, we derive $I_{\text{irs}}(\mathbf{r}_r)$ and the corresponding  channel gain $h_{\text{irs}}$. 
\subsection{ Reflected Electric Field }
First, in the following lemma, we determine the incident  electric field on the IRS plane.
 \begin{lem}\label{Lemma1}
 	Assuming that $d_i\gg L_x, L_y$, the  electric field  emitted by the LS incident on  the IRS plane, denoted by $E_i(\mathbf{r})$,  is given by 
 	\begin{IEEEeqnarray}{rll}
 	E_i(\mathbf{r})&= \frac{E_0 w_0}{{w}(\tilde{d}_i)}\exp\left(-{\hat{x}^2+{\hat{y}}^2\over {w}^2(\tilde{d}_i)}-j\phi_{in}\right)\quad\text{with phase}\nonumber\\
 			\phi_{in}&=k\left(\hat{d}_i+{\hat{x}^2+\hat{y}^2\over 2R(\tilde{d}_i)}\right)-\tan^{-1}\left(\frac{\tilde{d}_i}{z_0}\right),\quad
\label{lem1}
 \end{IEEEeqnarray} 
where $\mathbf{r}=(x, y, 0)$ denotes any point in the $xy$-plane, $\hat{d_i}=d_i+(x-x_{0})\cos(\theta_i)$, $\tilde{d}_i=d_i-x_{0}\cos(\theta_i)$,  $\hat{x}=\sin(\theta_i)(x-x_{0})$, and $\hat{y}=y-y_{0}$.	
 \end{lem}
\begin{IEEEproof}
The proof is given in Appendix \ref{App0}.
\end{IEEEproof}
Eq.~(\ref{lem1}) determines an elliptical Gaussian beam on the IRS with beam widths $w_x=\frac{w(\tilde{d}_i)}{\sin(\theta_i)}$ and $w_y=w(\tilde{d}_i)$ along $x$- and $y$-axes. 

Next,   to determine the impact of the IRS on the incident beam, we use scalar field theory \cite{Goodmanbook} and neglect the vectorial nature of the electromagnetic field. This approach yields  accurate result if the following conditions are met: 1) the diffracting surface must be large compared to the wavelength, 2) the electromagnetic fields must not be observed very close to the surface, i.e., $d_r\gg \lambda$ \cite{Goodmanbook}. Given the size of the IRS and the application of   FSO systems for long-distance communications, these conditions are met in practice. Thus, we can apply the Huygens-Fresnel principle for deriving  the  beam reflected by the IRS. This principle states that  every point on the wavefront of the beam can be considered as a secondary source emitting a  spherical wave and, at any position, the new wavefront  is determined by the sum of these  secondary waves \cite{Goodmanbook}.  Given this principle, the complex amplitude of the electric field  reflected by the IRS, denoted by $E_{r}(\mathbf{r}_{p})$, at an arbitrary observation point $\mathbf{r}_p=(x_p, y_p, z_p)$, see Fig.~\ref{Fig:System},   is given by \cite{Goodmanbook}
\iftoggle{Conf}{%
\begin{IEEEeqnarray}{rll}
	E_{r}(\mathbf{r}_{p})&=\frac{1}{j\lambda}\iint\limits_{\Sigma} \!E_i(\mathbf{r}) {e^{jk|\mathbf{r}_{op}|}\over |\mathbf{r}_{op}|} e^{j\Phi_\text{irs}(x, y)}  \cos({\theta}_{op})  \mathrm{d}x\mathrm{d}y,\,\quad
	\label{Huygens-Fresnel}
\end{IEEEeqnarray}
}{
\begin{IEEEeqnarray}{rll}
	E_{r}(\mathbf{r}_{p})&=\frac{1}{j\lambda}\iint_{\Sigma} E_i(\mathbf{r}) {\exp(jk|\mathbf{r}_{op}|)\over |\mathbf{r}_{op}|} \cos({\theta}_{op})  e^{j\Phi_\text{irs}(x, y)} \mathrm{d}x\mathrm{d}y,\,\quad
	\label{Huygens-Fresnel}
\end{IEEEeqnarray}
}
where   $\mathbf{r}_{op}$ is the vector between an arbitrary point on the IRS, denoted by vector $\mathbf{r}$, and any  observation point $\mathbf{r}_p$, ${\theta}_{op}$ denotes the angle between vector $\mathbf{r}_{op}$ and the $z$-axis, $\Sigma$  is   the IRS area, and  $\Phi_{\text{irs}}(x, y)$ denotes the  phase shift introduced by the IRS. In (\ref{Huygens-Fresnel}), the total surface of the IRS is divided into infinitesimal parts with area $\mathrm{d}x\mathrm{d}y$, and the light wave scattered by each  part is modeled as a secondary source emitting a spherical wave, modeled by ${\exp(jk|\mathbf{r}_{op}|)\over |\mathbf{r}_{op}|}$. The complex amplitudes of the secondary sources are proportional to the incident electric field, $E_i(\mathbf{r})$, and an additional phase shift term, $e^{j\Phi_\text{irs}(x, y)}$, is introduced by the IRS.  
The phases of the spherical sources, $k|\mathbf{r}_{op}|$, play an important role in our analysis, and for tractability, we approximate $|\mathbf{r}_{op}|$. The length of  vector $\mathbf{r}_{op}$, see Fig.~\ref{Fig:System}, is given by  
\begin{IEEEeqnarray}{rll}
|\mathbf{r}_{op}|&=|\mathbf{r}_p-\mathbf{r}|=\left[(x-x_p)^2+(y-y_p)^2+z_p^2\right]^{1/2}.
\end{IEEEeqnarray}	
Let  $d_p=|\mathbf{r}_p|$, then we obtain
\begin{IEEEeqnarray}{rll}	
	&\frac{|\mathbf{r}_{op}|^2}{d_p^2}=1+\frac{x^2+y^2}{d_p^2}-2\frac{xx_p+yy_p}{d_p^2}.\qquad
\end{IEEEeqnarray}
Applying the Taylor series expansion \cite{integral} with $(1+x)^{1/2}=1+\frac{1}{2} x-\frac{1}{8} x^2+ \cdots$, we obtain
\iftoggle{Conf}{%
\begin{IEEEeqnarray}{rll}	
	{|\mathbf{r}_{op}|}&=\underbrace{d_p-{\frac{xx_p+yy_p}{d_p}}}_{=\text{t}_1}+\underbrace{\frac{x^2+y^2}{2d_p}}_{=\text{t}_2}\nonumber\\
		&+\underbrace{\frac{(x^2+y^2-2(xx_p+yy_p))^2}{8d_p^3}}_{=\text{t}_3}+\cdots
	.\qquad
	\label{bionom}
\end{IEEEeqnarray}
}{
\begin{IEEEeqnarray}{rll}	
	&{|\mathbf{r}_{op}|}=\underbrace{d_p-{\frac{xx_p+yy_p}{d_p}}}_{=\text{t}_1}+\underbrace{\frac{x^2+y^2}{2d_p}}_{=\text{t}_2}+\underbrace{\frac{(x^2+y^2-2(xx_p+yy_p))^2}{8d_p^3}}_{=\text{t}_3}+\cdots
	.\qquad
	\label{bionom}
\end{IEEEeqnarray}
}
For the far-field approximation, it is assumed that the rays reflected by the IRS surface are parallel to each other and the rays have only a linear phase shift w.r.t. each other \cite{Hecht}. This approximation is equivalent  to  assuming a linear phase shift for the phase of the secondary sources w.r.t.  the $x$- and $y$-directions. In other words, only $\text{t}_1$ in (\ref{bionom}) is taken into account, and     $\text{t}_2$ and all higher orders terms are  neglected. For the far-field assumption to hold,  the impact of   $\text{t}_2$ in the argument of the  exponential term,  $k|\mathbf{r}_{op}|$, should be much smaller than one period of the complex exponential, and thus, 
  \begin{IEEEeqnarray}{rll}	
k\frac{x^2+y^2}{2d_p}\ll 2\pi.
\label{phass}
\end{IEEEeqnarray}
The range of the  relevant values for $x$ and $y$ in  (\ref{Huygens-Fresnel}) is bounded by the beam widths of the incident electric field $2w_x$ and $2w_y$ (where the power of the incident beam drops by $\frac{1}{e^4}$ compared to the peak value)    and the size of the IRS $L_x$ and $L_y$, i.e., $x_e=\min\left(\frac{L_x}{2}, w_x\right)\geq |x|$ and $y_e=\min\left(\frac{L_y}{2}, w_y\right)\geq |y|$. For practical lens and IRS sizes, we have $a_0, L_x, L_y\ll d_r$, and hence, $d_p\approx d_r$ holds, see Fig.~\ref{Fig:System}. Thus,  substituting $x_e$ and $y_e$ for $x$ and $y$ in (\ref{phass}), respectively,  and  defining the far-field distance as follows
 \begin{IEEEeqnarray}{rll}	
d_f=\frac{x_e^2+y_e^2}{8\lambda},
\label{d_f}
\end{IEEEeqnarray}
  for  distances $d_r\approx d_p \gg d_f$, the approximation of (\ref{bionom}) in (\ref{Huygens-Fresnel}) by only term $\text{t}_1$ is appropriate.  However, depending  on the values of $x_e$ and $y_e$, and the observation distance  from the IRS, $d_r$, $d_r\gg d_f$ might not hold. For example, consider a typical IRS size  of $L_x=L_y=50$ cm and an FSO beam with wavelength of  $\lambda=1550$ nm illuminated from a LS at $\mathbf{p}_i= \left(d_i,\theta_i, \phi_i\right)=\left(1000\, \text{m}, \frac{\pi}{8}, 0\right)$. Then, the beam incident on the IRS has widths  $w_x=0.52$ m and $w_y=0.19$ m and the far-field distance according to (\ref{d_f}) is $d_f=32.7$ km  and the observation point should be farther away from the center of the IRS  than this distance, which is not practical. Thus, in order to  obtain a model that is also valid for   intermediate distances, we propose the following theorem which is  valid for both intermediate and far-field distances. Assuming a linear phase shift profile across the IRS, a closed-form solution for the integral in (\ref{Huygens-Fresnel}) is determined in this theorem.
\begin{theo}\label{Theorem1}
Assuming a linear phase-shift profile across the IRS, i.e.,  $\Phi_\text{irs}(x,y)\!=\!k\left(\!\Phi_x x+\Phi_y y\!\right)$, where $\Phi_x$ and $\Phi_y$ are the constant phase-shift gradients in  $x$- and $y$-direction, respectively, then the  electric field emitted by the LS  at position $\mathbf{p}_i=(d_i, \theta_i, \phi_i)$  and reflected by the IRS  at the lens located at $\mathbf{p}_r=(d_r, \theta_r, \phi_r)$   for any distance $d_i\gg L_x,L_y$, $d_r \gg a_0, {L_x}, L_y, d_n$, where 
\begin{IEEEeqnarray}{rll}
d_n=\left[\frac{(x_e^2+y_e^2)^2}{8\lambda}\right]^{1/3},
\end{IEEEeqnarray}
 is given by  (\ref{theo1}), shown on top of the next page.  In  (\ref{theo1}), we use  $X=\varphi_x+c_1x_r+c_2y_r+2\hat{b}_{x}x_{0}$, $Y=\varphi_y+c_3x_r+c_4y_r+2\hat{b}_{y}y_{0}$,
$C= \frac{E_0w_0}{j\lambda{w}(\tilde{d}_i)d_r}e^{jk(-\tilde{d_i}+d_r)+jkx_{0}\cos(\theta_i)+j\!\tan^{\!\!\!-1}\!\!\left(\frac{\tilde{d}_i}{z_0}\right)-\nu\sin^{\!2}\!(\theta_i) x^2_{0}-\nu y^2_{0}}$,   $c_1 = \frac{1}{d_r}\cos(\phi_r)\sin(\theta_r)$, $c_2=\frac{1}{d_r}\sin(\phi_r)$, $c_3=\frac{-1}{d_r}\sin(\phi_r)\sin(\theta_r)$, $c_4=\frac{1}{d_r}\cos(\phi_r)$,  $\varphi_x=\cos(\theta_i)+\cos(\theta_r)\cos(\phi_r)$,  $\varphi_y=-\cos(\theta_r)\sin(\phi_r)$, $b_x=\nu\sin^2(\theta_i)-\frac{jk}{2d_r}\left(1+\cos^2(\phi_r)\cos^2(\theta_r)\right)$, $b_y=\nu-\frac{jk}{2d_r}\left(1-\sin^2(\phi_r)\cos^2(\theta_r)\right)$, $\nu={1\over w^2(\tilde{d}_i)}+{jk\over 2R(\tilde{d}_i)}$, $\hat{b}_{x}=-j\frac{b_{x}}{k}$, and  $\hat{b}_{y}=-j\frac{b_{y}}{k}$.  
\begin{figure*}
		\begin{IEEEeqnarray}{rll}
	&E_{r}(\mathbf{r}_r)={C} {\frac{{\pi\sin\theta_r}}{4\sqrt{b_x b_y}}}\, e^{-\frac{k^2}{4b_x}\left(X-\Phi_x\right)^2-\frac{k^2}{4b_y}\left(Y-\Phi_y\right)^2}\nonumber\\
	&\times\Bigg[\text{erf}\left({\sqrt{b_x}}\frac{L_x}{2}-\frac{jk}{2\sqrt{b_x}}\left(X-\Phi_x\right)\right)-\text{erf}\left(-{\sqrt{b_x}}\frac{L_x}{2}-\frac{jk}{2\sqrt{b_x}}\left(X-\Phi_x\right)\right)\Bigg]\nonumber\\
	&\times\Bigg[\text{erf}\left({\sqrt{b_y}}\frac{L_y}{2}-\frac{jk}{2\sqrt{b_y}}\left(Y-\Phi_y\right)\right)-\text{erf}\left(-{\sqrt{b_y}}\frac{L_y}{2}-\frac{jk}{2\sqrt{b_y}}\left(Y-\Phi_y\right)\right)\Bigg]\quad
	\label{theo1}
\end{IEEEeqnarray}
	\noindent\rule{\textwidth}{0.8pt}
\end{figure*}
\end{theo}
\begin{IEEEproof}
The proof is given in Appendix \ref{App1}.
\end{IEEEproof}
Eq.~(\ref{theo1})  explicitly shows the impact of the positioning of the LS and the lens w.r.t. the IRS,  the size of the IRS, and  the phase-shift configuration across the IRS on the electric field reflected by the IRS. In contrast to previous results for far-field approximations, the above theorem is valid even for intermediate distances. For the previous example,  we obtain $d_n=9.4$ m and for IRS-lens distances, $d_r$, larger than $d_n$, the result in (\ref{theo1}) is  accurate.  We may adopt  the linear phase shifts proposed in previous works \cite{Vahid_IRS} to configure the IRS.   In this case, $\Phi_x$ and $\Phi_y$ are set  as follows \cite{Vahid_IRS}
\begin{IEEEeqnarray}{rll}
	\Phi_x&=\cos(\theta_i)\cos(\phi_i)+\cos(\theta_r)\cos(\phi_r), \nonumber\\ 	
	\Phi_y&=\cos(\theta_i)\sin(\phi_i)+\cos(\theta_r)\sin(\phi_r).
	\label{LP}
\end{IEEEeqnarray}
In the following corollary, we consider a special case of Theorem \ref{Theorem1} by assuming the conventional mirror. A conventional mirror introduces  no additional phase shifts, i.e., $\Phi_x=\Phi_y=0$, and the incident angle and the reflection angle follow Snell's law, i.e., $\theta_i=\theta_r$. 
\begin{corol}[Reflection by Conventional Mirror]\label{col1}
Assume the  size of the conventional mirror is very large, i.e., ${L_x}, L_y\gg 2w(\tilde{d_i})$, such that  the entire received beam is reflected.  Then, assuming a  far-field scenario, $d_r\gg d_f$, $R(\tilde{d}_i) \to \infty$, $\phi_r=\pi-\phi_i$ and $x_0=y_0=0$,  (\ref{theo1}) simplifies to	
	\begin{IEEEeqnarray}{rll}
		E_{r}(\mathbf{r}_r)&=C{\pi}{w^2(\tilde{d}_i)} \exp\left(-\frac{x_r^2+y_r^2}{w_{eq}^2}\right),
		\label{GEo-Gauss}
	\end{IEEEeqnarray}
	where $w_{eq}=\frac{2d_r}{k w(\tilde{d}_i)}$. 
\end{corol}
\begin{IEEEproof}
Considering ${L_x}, L_y\gg 2w(\tilde{d_i})$, we can substitute the $\text{erf}(\cdot)$ terms in (\ref{theo1}) by 4. Assuming $R(\tilde{d}_i) \to \infty$ and $d_r\gg d_f$, we obtain $b_x=\frac{\sin^2(\theta_i)}{w^2(\tilde{d}_i)}$ and $b_y=\frac{1}{w^2(\tilde{d}_i)}$. Substituting	$\theta_i=\theta_r$, $\phi_r=\pi$, and $\Phi_x=\Phi_y=0$, leads to (\ref{GEo-Gauss}) and this completes the proof.
\end{IEEEproof}
Eq.~(\ref{GEo-Gauss}) corresponds to a circular Gaussian beam, and reveals that in the assumed regime the reflected beam is identical to what  is expected from geometrical optics under the far-field assumption, see \cite{Marzieh_IRS}.  
\begin{corol}[Reflection by Anomalous Mirror]\label{col2}
	Consider an anomalous mirror which  imposes the additional phase shifts, $\Phi_x$ and $\Phi_y$, given in (\ref{LP}).  Then, for the far-field scenario, $d_r\gg d_f$, $R(\tilde{d}_i) \to \infty$,  ${L_x}, L_y\gg 2w(\tilde{d_i})$, and  $x_0=y_0=0$,  (\ref{theo1}) simplifies to
\iftoggle{Conf}{%
	\begin{IEEEeqnarray}{rll}
		E_{r}(\mathbf{r}_r)&= {\frac{C\pi|\sin(\theta_r)| w^2(d_i)}{|\sin(\theta_i)|}} \nonumber\\
		 &\times\exp\left(-\frac{k^2 w^2(d_i)}{4d_r^2}\left(\frac{\sin^2(\theta_r)x_r^2}{\sin^2(\theta_i)}+y_r^2\right)\right).
		\label{ellips_Gaus}
	\end{IEEEeqnarray}
}{
\begin{IEEEeqnarray}{rll}
E_{r}(\mathbf{r}_r)&= {\frac{C\pi|\sin(\theta_r)| w^2(d_i)}{|\sin(\theta_i)|}} \exp\left(-\frac{k^2 w^2(d_i)}{4d_r^2}\left(\frac{\sin^2(\theta_r)x_r^2}{\sin^2(\theta_i)}+y_r^2\right)\right).
\label{ellips_Gaus}
\end{IEEEeqnarray}
}
\end{corol}
\begin{IEEEproof}
	Substituting $\Phi_x$ and  $\Phi_y$, given by (\ref{LP}),   compensates for $\varphi_x$ and $\varphi_y$ in (\ref{theo1}), respectively. Then, considering $\theta_i\neq\theta_r$ and following similar steps as in the proof of Corollary \ref{col1}  leads to (\ref{ellips_Gaus}). This completes the proof. 
\end{IEEEproof}
Eq.~(\ref{ellips_Gaus}) describes an elliptical Gaussian beam. This result is  in agreement with the result from geometric optics for far-field reflection by an IRS  \cite{Marzieh_IRS_jou}.

Depending on the IRS size,   distances $d_r\gg d_f$ might not be in the practical range    for  FSO systems, see Section \ref{Sec_Sim}. Therefore,  Corollaries \ref{col1} and \ref{col2},  which are in-line with the far-field approximation,  do not always provide a valid result.  Thus, Theorem \ref{Theorem1}  is required to determine the channel gain for  practical applications, see Section \ref{Sec_Sim}.
 
\subsection{Deterministic Channel Gain of the IRS}
 Given (\ref{theo1}),  the power distribution, $I_\text{irs}(\mathbf{r}_r)$,  and  the corresponding deterministic channel gain, $h_{\text{irs}}$, can be obtained with (\ref{irradiance}) and (\ref{gain}), respectively. In the following theorem, we assume $d_r\gg d_n$ for the distance of the lens from the IRS and  provide the deterministic channel gain.
\begin{theo}[Out-of-Plane Reflection]\label{Theorem4}
	Assume $d_i\gg L_x,L_y$, $d_r\gg a_0, {L_x}, L_y, d_n$,  a linear phase-shift configuration of the IRS,  $\Phi_\text{irs}=k\left(\Phi_x x+\Phi_y y\right)$, and a LS located at $\mathbf{p}_i$. Then, the  channel gain for the lens at position $\mathbf{p}_r$ is given by
\iftoggle{Conf}{%
\begin{IEEEeqnarray}{rll}
	& 	h^{\text{out}}_{\text{irs}}=\frac{{C}_h\sqrt{\pi}}{2\sqrt{\rho_x}}\int\limits_{-\varepsilon}^{\varepsilon}  \exp\left(-\rho_yy_r^2-\varrho_{y}y_r+\frac{(\rho_{xy}y_r+\varrho_{x})^2}{4\rho_x}\right)\times\nonumber\\ 
	&\Bigg[\text{erf}\left(\!\!\frac{\rho_{xy} y_r+{2\rho_x\varepsilon} +\varrho_{x}}{2\sqrt{\rho_x}}\!\!\right)-\text{erf}\left(\frac{{\rho_{xy} y_r-2\rho_x\varepsilon} +\varrho_{x}}{2\sqrt{\rho_x}}\right)\Bigg] \mathrm{d}y_r,\nonumber\\
	& \qquad
			\label{theo4}
\end{IEEEeqnarray}
	}{	
\begin{IEEEeqnarray}{rll}
		h^{\text{out}}_{\text{irs}}=\frac{{C}_h}{2} \sqrt{\frac{\pi}{\rho_x}}& \int\limits_{-\varepsilon}^{\varepsilon}   \exp\left(-\rho_yy_r^2-\varrho_{y}y_r+\frac{(\rho_{xy}y_r+\varrho_{x})^2}{4\rho_x}\right)\nonumber\\ &\times\Big[\text{erf}\left(\frac{{2\rho_x}+\rho_{xy} y_r +\varrho_{x}}{2\sqrt{\rho_x}}\right)\!-\text{erf}\left(\frac{\rho_{xy} y_r -2\rho_x+\varrho_{x}}{2\sqrt{\rho_x}}\right)\Big] \mathrm{d}y_r,\qquad
		\label{theo4}
\end{IEEEeqnarray}
}
\iftoggle{Conf}{%
where $\varepsilon=\frac{\sqrt{\pi}a_0}{2}$,  $\tilde{b}_i={2b_i\bar{b}_i\over b_i+\bar{b}_i}$, $\bar{b}_i=\text{conj}(b_i)$, $i\in \{x, y\}$,  $\rho_1=\varrho_y+\varepsilon\rho_{xy}$, $\rho_2= \varrho_y-\varepsilon\rho_{xy}$, $\!C_h=\frac{\pi^2{|C_2|^2}E_0^2 {w_0^2}\sin^2(\theta_r)}{32\eta{|b_x|} {|b_y|}{\lambda^2{w}^2(\tilde{d}_i)d_r^2}}\times
\exp\left({{(\varphi_x-\Phi_x)^2\over \tilde{b}_x}+{\left(\varphi_y-\Phi_y\right)^2\over \tilde{b}_y}-2\nu\left(\sin^2(\theta_i)x_0^2+ y_0^2\right)-\frac{k^2}{2}}\right)$, $\rho_x= \frac{k^2}{2}\left(\frac{c_1^2}{\tilde{b}_x}+\frac{c_3^2}{\tilde{b}_y}\right)$, $\rho_y=\frac{k^2}{2} \left(\frac{c_2^2}{\tilde{b}_x}+\frac{c_4^2}{\tilde{b}_y}\right)$, $\varrho_{x}={k^2}\left({c_1\over \tilde{b}_x}\left(\varphi_x-\Phi_x\right)+{c_3\over \tilde{b}_y}\left(\varphi_y-\Phi_y\right)\right)$, $\varrho_{y}={{k^2}c_4\over \tilde{b}_y}\left(\varphi_y-\Phi_y\right)-{{k^2}c_2\over \tilde{b}_x}\left(\varphi_x-\Phi_x\right)$, $\rho_{xy}=\frac{k^2c_1c_2}{\tilde{b}_x}+\frac{k^2c_3c_4}{\tilde{b}_y}$,     $\varpi_x= -\sqrt{b_x}\frac{{L}_{x}}{2}-\frac{jk}{2\sqrt{b_x}}\left(\varphi_x-\Phi_x+2\hat{b}_xx_{0}+(c_1+c_2)a_0\right)$, $\varpi_y= -\sqrt{b_y}\frac{{L}_{y}}{2}-\frac{jk}{2\sqrt{b_y}}\left(\varphi_y-\Phi_y+2\hat{b}_yy_{0}+(c_3+c_4)a_0\right)$, 
$C_2\!=\left(\text{erf}\left(\sqrt{b_x}{L}_{x}+\varpi_x\right)\!\!-\!\!\text{erf}\left(\varpi_x \right)\right)\!\!\left(\text{erf}\left(\sqrt{b_y}{L}_{y}+\varpi_y\right)\!\!-\!\!\text{erf}\left(\varpi_y\right)\right)$.	
}{
where $\varepsilon=\frac{\sqrt{\pi}a_0}{2}$, $\tilde{b_i}={2b_i\bar{b}_i\over b_i+\bar{b}_i},\bar{b}_i=\text{conj}(b_i),$ $i\in \{x, y\}$, $\rho_1=\varrho_y+\varepsilon\rho_{xy}$, $\rho_2= \varrho_y-\varepsilon\rho_{xy}$,  $\rho_x= \frac{k^2}{2}\left(\frac{c_1^2}{\tilde{b}_x}+\frac{c_3^2}{\tilde{b}_y}\right)$, $\rho_y=\frac{k^2}{2} \left(\frac{c_2^2}{\tilde{b}_x}+\frac{c_4^2}{\tilde{b}_y}\right)$, $\rho_{xy}=k^2 \left(\frac{c_1c_2}{\tilde{b}_x}+\frac{c_3c_4}{\tilde{b}_y}\right)$, $\varrho_{x}={k^2}\left({c_1\over \tilde{b}_x}\left(\varphi_x-\Phi_x\right)+{c_3\over \tilde{b}_y}\left(\varphi_y-\Phi_y\right)\right)$,  $\varrho_{y}={k^2}\left(-{c_2\over \tilde{b}_x}\left(\varphi_x-\Phi_x\right)+{c_4\over \tilde{b}_y}\left(\varphi_y-\Phi_y\right)\right)$, $\varpi_x= -\sqrt{b_x}\frac{{L}_{x}}{2}-\frac{jk}{2\sqrt{b_x}}\left(\varphi_x-\Phi_x+2\hat{b}_xx_{0}+(c_1+c_2)a\right)$, $\varpi_y= -\sqrt{b_y}\frac{{L}_{y}}{2}-\frac{jk}{2\sqrt{b_y}}\left(\varphi_y-\Phi_y+2\hat{b}_yy_{0}+(c_3+c_4)a\right)$, 
$C_2=\left(\text{erf}\left(\sqrt{b_x}{L}_{x}+\varpi_x\right)-\text{erf}\left(\varpi_x \right)\right)\times$ $\left(\text{erf}\left(\sqrt{b_y}{L}_{y}+\varpi_y\right)-\text{erf}\left(\varpi_y\right)\right)$,  $C_h=\frac{\pi^2{|C_2|^2}E_0^2 {w_0^2}\sin^2(\theta_r)}{32\eta{|b_x|} {|b_y|}{\lambda^2{w}^2(\tilde{d}_i)d_r^2}}
e^{-2\nu\left(\sin^2(\theta_i)x_0^2+ y_0^2\right)}e^{-\frac{k^2}{2}\left({\left(\varphi_x-\Phi_x\right)^2\over \tilde{b}_x}+{\left(\varphi_y-\Phi_y\right)^2\over \tilde{b}_y}\right)}$.}
\end{theo}
\begin{IEEEproof}
The proof is given in Appendix \ref{App4}.
\end{IEEEproof}
Theorem \ref{Theorem4} specifies  the channel gain of an IRS-assisted FSO system when the normal vector of the LS plane and the normal vector of the lens plane may lie in different  planes, which is referred to as ``out-of-plane reflection'' \cite{outofreflection}.  This is  in contrast to Snell's law which states that the reflected and incident beams should be in the same plane with $\theta_i=\theta_r$. However, by adopting a linear phase shift at the IRS, the  direction of the reflected beam can be out of the incident beam plane.
 Moreover, (\ref{theo4}) specifies the dependence of the channel gain on the IRS size, $L_x$ and $L_y$,  the phase gradients across the IRS surface, $\Phi_x$ and $\Phi_y$, the lens radius, $a_0$, and the lens and LS positions, $\mathbf{p}_r$ and $\mathbf{p}_i$, respectively.
Furthermore,  the position of the center of the beam footprint on the IRS surface, $x_0$ and $y_0$, or in other words the position where the Gaussian beam intersects the IRS surface, affects the channel gain. This is expected since when the beam is not accurately tracked, only a fraction of total power is received by the lens which degrades the channel gain.

In the following corollary, we simplify (\ref{theo4}), for the case where the normal vector of the LS plane and the normal vector of the lens plane lie in the same plane, which is referred to as ``in-plane-reflection'' \cite{outofreflection}.
\begin{corol}[In-Plane Reflection]
 For in-plane reflection,  $\phi_r=\pi-\phi_i$, and the channel gain   is given by
	\begin{IEEEeqnarray}{rll}
		h^\text{in}_\text{irs}&= \frac{C_h \pi}{4\sqrt{\rho_x\rho_y}} \exp\left(\frac{\varrho_x^2}{4\rho_x}+\frac{\varrho_y^2}{4\rho_y}\right)\nonumber\\ 
		&\times \left[\text{erf}\left(\sqrt{\rho_x}\varepsilon+\frac{\varrho_{x}}{2\sqrt{\rho_x}}\right)-\text{erf}\left(-\sqrt{\rho_x}\varepsilon+\frac{\varrho_{x}}{2\sqrt{\rho_x}}\right)\right]\nonumber\\
		&\times \left[\text{erf}\left(\sqrt{\rho_y}\varepsilon+\frac{\varrho_{y}}{2\sqrt{\rho_y}}\right)-\text{erf}\left(-\sqrt{\rho_y}\varepsilon+\frac{\varrho_{y}}{2\sqrt{\rho_y}}\right)\right]\!.\! \qquad
		\label{col}
	\end{IEEEeqnarray}
\end{corol}
\begin{IEEEproof}
	 Since $\phi_i=0$ and   $\phi_r=\pi$,  the parameters in Theorem \ref{Theorem4}  simplify to	 $\rho_x=-\frac{k^2}{2d_r^2\tilde{b}_x}\sin^2(\theta_r)$, $\rho_y=-\frac{k^2}{2d_r^2\tilde{b}_y}$,  $\rho_{xy}=0$, $\varrho_{x}=\frac{k^2}{d_r \hat{b}_x}\sin(\theta_r)\left(\varphi_x-\Phi_x\right)$, $\varrho_{y}=0$, $\varphi_y=0$, and $\varphi_x=-\cos(\theta_r)+\cos(\theta_i)$. Substituting these values, the integral in   (\ref{ptheo43})  simplifies  to	 two independent integrals, which can be solved by applying  \cite[Eq.~(2.33-1)]{integral}. Then, we obtain (\ref{col}) and this completes the proof.
\end{IEEEproof}

\subsection{BER Performance Analysis}
Assuming OOK modulation, the average BER, denoted by $P_e$,   over a Gamma-Gamma fading channel is given by \cite{Error-Robert}
\begin{IEEEeqnarray}{rll}
	  P_{e} =& \sum_{\ell=0}^{\infty}\left(\xi_{\ell}(\alpha, \beta)\gamma^{-{\ell+\beta\over 2}}+\xi_{\ell}(\beta, \alpha)\gamma^{-{\ell+\alpha\over 2}}\right),
	  \label{BER}
\end{IEEEeqnarray}
where 
\begin{IEEEeqnarray}{rll}
 &\xi_{\ell}(\alpha, \beta) =\nonumber\\&\quad{\sqrt{\pi}(2\sqrt{2}\alpha\beta)^{\ell+\beta}{\Gamma}\left({\ell+\beta+1\over 2}\right)\over 2\sin[\pi(\alpha-\beta)]\Gamma(\alpha)\Gamma(\beta) \Gamma(\ell-\alpha+\beta+1)(\ell+\beta)\ell!}. \qquad
\end{IEEEeqnarray}
Here, $\Gamma(\cdot)$ is the Gamma function and the signal-to-noise ratio term is denoted by $\gamma=\frac{\left(h_p h_\text{irs}\right)^2\sigma_s^2}{\sigma_n^2}$.

\section{Simulation Results}\label{Sec_Sim}
In this section,    we  validate  our analytical results for the electric field in (\ref{theo1}) and the channel gain of  IRS-assisted FSO systems in (\ref{theo4})  and  investigate  the BER performance. We consider a  square-shaped IRS with  $L_x=L_y=0.5$ m and adopt the linear phase configuration given in (\ref{LP}). The LS is positioned at  $\mathbf{p}_i=\left(d_i, \theta_i, \phi_i\right)=(1000\, \text{m}, \frac{\pi}{8}, 0)$ emitting a Gaussian beam with parameters  $\lambda=1550\ \mathrm{nm}$, $E_0=1000\ \frac{\mathrm{V}}{\mathrm{m}}$, and $w_0=2.5\ \mathrm{mm}$. The  receiver lens is located at $\mathbf{p}_r=\left(d_r, \theta_r, \phi_r\right)=(2000\, \text{m}, \frac{\pi}{2}, \pi)$. We adopt for the impedance of the propagation medium $\eta=377 \, \Omega$, for the Gamma-Gamma turbulence parameters $\alpha=2.1$ and $\beta=2$, and for the attenuation factor $\kappa=16.8\times 10^{-3} \frac{\text{dB}}{\text{m}}$.

\begin{figure}
	\centering
	\includegraphics[width=0.55\textwidth]{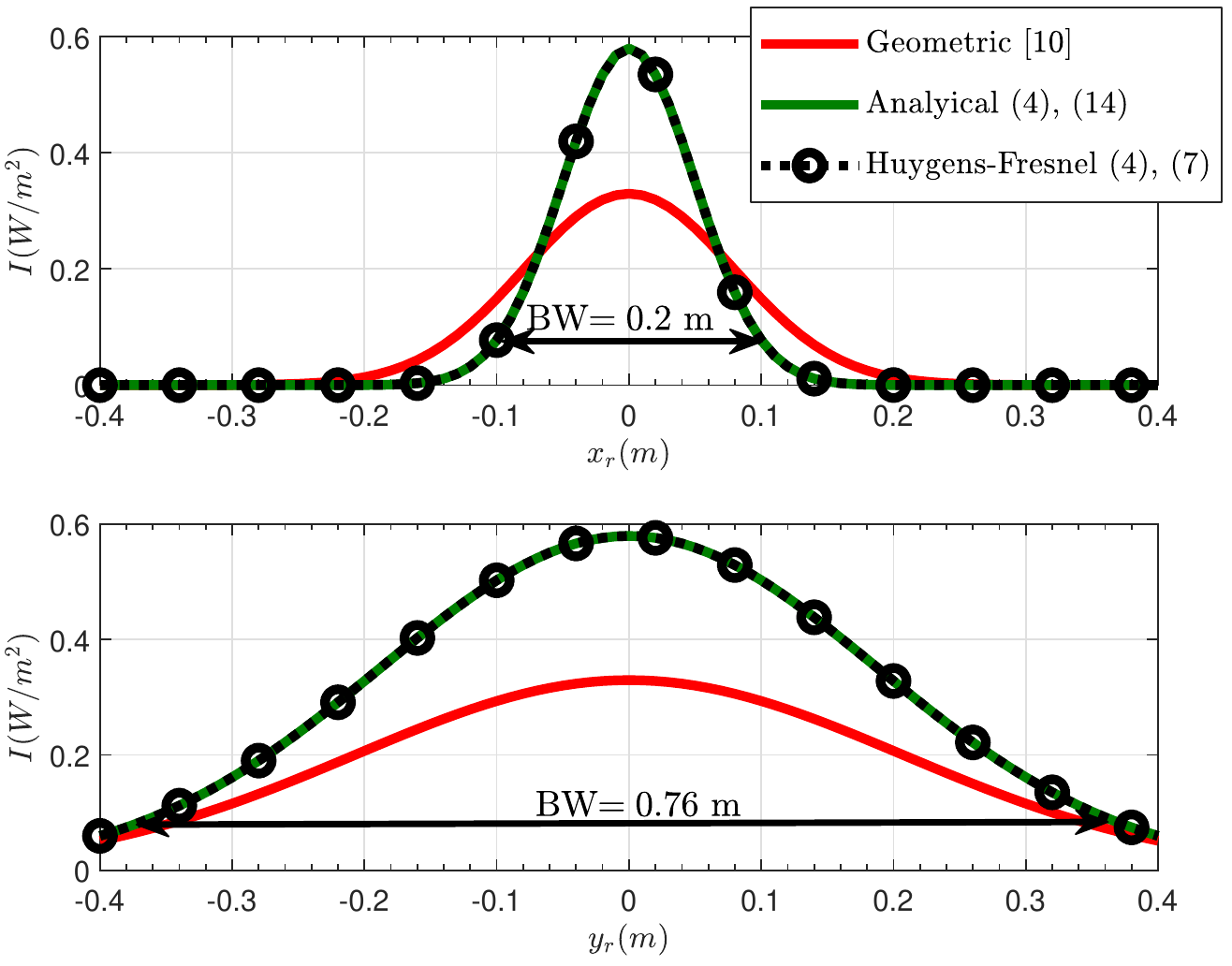}
	\caption{Power density in the lens plane versus $x_r$- and $y_r$.}
	\label{Fig:near}\vspace{-0.3cm}
\end{figure}

The top and bottom subfigures of Fig.~\ref{Fig:near} show the power density (\ref{irradiance}) of the  Gaussian beam   along the $x_r$- and $y_r$-axes of the receiver lens coordinate system, respectively. Results for the power density according to our  model in  (\ref{theo1}),  numerical integration according to the Huygens-Fresnel principle using   (\ref{Huygens-Fresnel}), which serves as ground truth,   and   geometric optics under the far-field approximation    \cite{Marzieh_IRS_jou} are shown.
Since for the beam width of the incident laser beam at the IRS, $w_x=0.52\,\text{m}\simeq L_x$  and $w_y=0.19 \,\text{m} \leq L_y$  holds,  the IRS can  reflect most of the power it receives to the lens. As can be observed from Fig.~\ref{Fig:near}, for the considered case, the result for the  far-field approximation in   \cite{Marzieh_IRS_jou} does not match the Huygens-Fresnel result. This is expected from our analysis since   $d_r=2000$ m  is much smaller than the far-field distance, $d_f=32.7$ km, see (\ref{d_f}). However, since the proposed model in  (\ref{theo1}) is valid for distances $d_r\gg d_n=9.4$ m, the power density obtained with this model perfectly matches the Huygens-Fresnel result. This confirms that, for practical  IRS sizes and realistic IRS-lens distances, the proposed  model has to be used  to accurately model IRS-assisted FSO channels. Moreover, we observe in Fig.~\ref{Fig:near} that the beam on the lens plane  is an elliptical Gaussian beam   with  beam widths 0.2 m and 0.76 m in the $x_r$- and $y_r$ directions, respectively. While the IRS  can adjust its phase-shift configuration such that the beam originating from  the LS  is reflected towards the lens,  the circular shape of the laser beam changes to an ellipse on the lens.   
\begin{figure}
	\centering
	\includegraphics[width=0.55\textwidth]{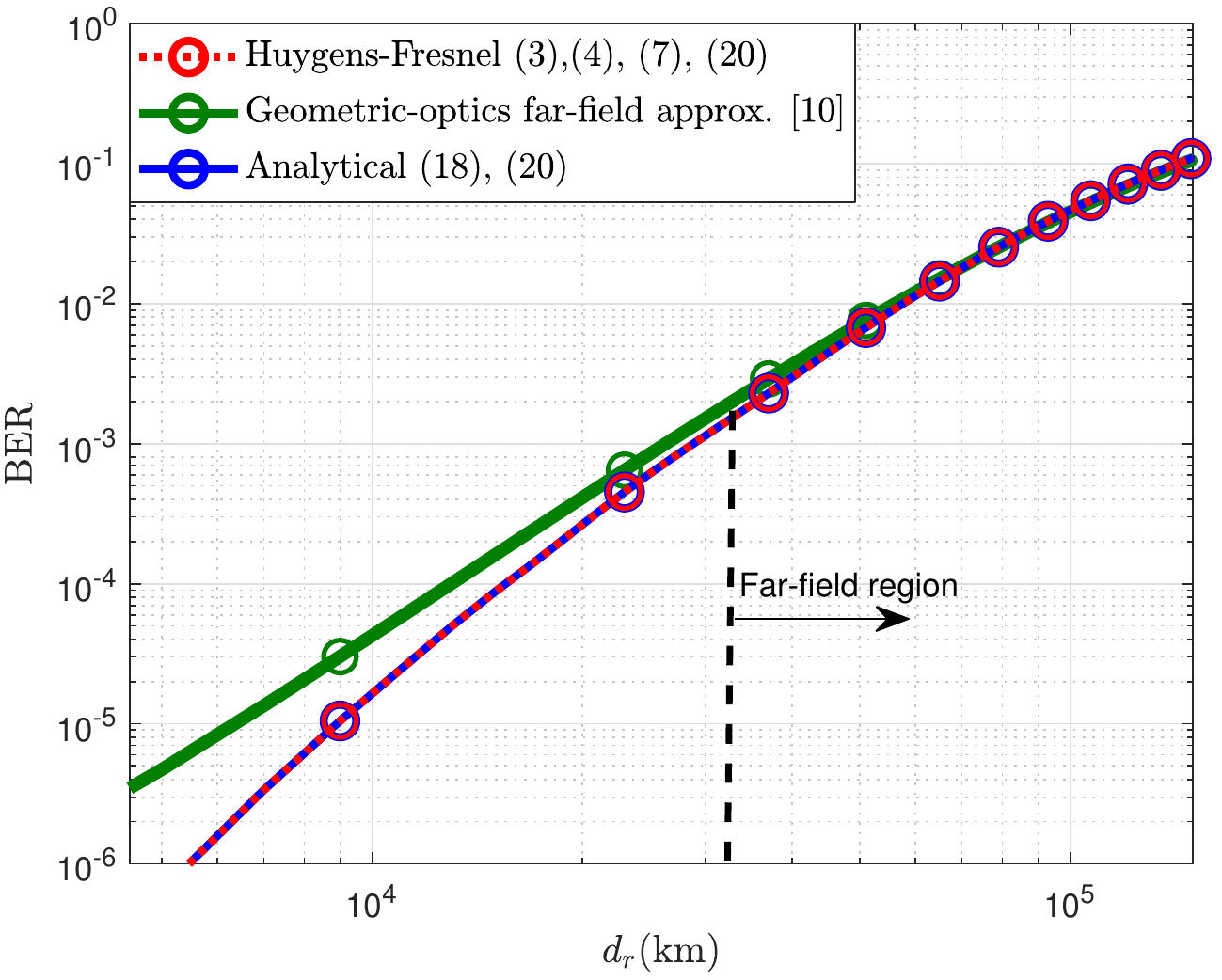}
	\caption{BER performance versus $d_r$.}
	\label{Fig:BER}\vspace{-0.3cm}
\end{figure}

Fig.~\ref{Fig:BER} shows the BER performance as obtained with (\ref{BER}) for the considered setup versus  the distance of the  lens center, $d_r$, from the IRS. We validate the proposed model for the channel gain $h_\text{irs}$  (\ref{theo4}) needed in (\ref{BER}) with numerical integration of the Huygens-Fresnel principle using  (\ref{gain}),  (\ref{irradiance}), and (\ref{Huygens-Fresnel}) and compare it with  results for the geometric optics model with the far-field approximation in  \cite{Marzieh_IRS_jou}. As  can be observed, by increasing the IRS-lens distance, $d_r$, the BER performance degrades due to increased geometric and atmospheric losses. Moreover, the proposed model perfectly matches the results obtained with  the Huygens-Fresnel principle  for intermediate and far-field distances  since for the considered $d_r$, $d_r\gg d_n=9.4$ m holds.  However, the BER for the far-field approximation approaches the Huygens-Fresnel result only for exceedingly large distances, i.e., for  $d_r> d_f=32.7$ km.  This confirms again that for practical applications of FSO systems, the far-field approximation does not yield realistic results and the proposed model has to be used instead.

\section{Conclusions}\label{Sec_concl}
In this paper, we developed  an analytical  channel model for 	IRS-assisted FSO systems based on the Huygens-Fresnel principle. We determined the reflected electric field and the channel gain   taking into account  the non-uniform power distribution  of   Gaussian beams, the impact of the IRS size, the positions of the LS, the IRS, and the lens, and the phase-shift configuration of the IRS. We validated the accuracy of the proposed analytical model via simulations and showed that,  in contrast to a model based on geometric optics employing the far-field approximation, the proposed model is valid even for intermediate distances, which are relevant in practice.
      
\appendices
\renewcommand{\thesectiondis}[2]{\Alph{section}:}
\section{Proof of Lemma~\ref{Lemma1}}\label{App0}
First, we transform  the LS plane coordinates $\mathbf{r}_i$ to  Cartesian coordinates $\mathbf{r}$ as follows  
\begin{IEEEeqnarray}{rll}
{\mathbf{r}}_i=\left(\mathbf{R}_{y_i}\left({\frac{\pi}{2}}-\theta_i\right)\mathbf{R}_z(-{\phi_i})\right)^T \left(\mathbf{r}-\mathbf{r}_{0}\right)+(0,0,d_i),\quad
\label{proof_lem11}
\end{IEEEeqnarray}
where 
$
\mathbf{R}_z({\phi})=
\begin{pmatrix}
\cos(\phi) & \sin(\phi) &0\\
-\sin(\phi) & 	\cos(\phi) & 0\\
0 & 0 & 1
\end{pmatrix}
$ and $\mathbf{R}_y({\phi})=
\begin{pmatrix}
\cos(\phi) & 0 & -\sin(\phi)\\
0 & 1 & 0\\
\sin(\phi) & 0 & \cos(\phi)
\end{pmatrix}$ denote the counter-clockwise rotations by angle $\phi$  around the $z$- and $y$-axes, respectively. Then, without loss of generality, we assumed $\phi_i=0$, thus, $	\mathbf{R}_{\phi}=\mathbf{I}_3$. To project on the IRS plane, we substitute $z=0$ in (\ref{proof_lem11}). Then, assuming $d_i\gg L_x$, we can approximate $\hat{d}_i\approx \tilde{d}_i$ in the terms ${w}^2(\cdot)$, $R(\cdot)$, and $\tan^{-1}\left(\cdot\right)$  in (\ref{Gauss}).    Then, by substituting (\ref{proof_lem11}) in (\ref{Gauss}), we obtain (\ref{lem1}) and this completes the proof.

 \section{Proof of Theorem~\ref{Theorem1}}\label{App1}
First, we substitute (\ref{lem1}) into (\ref{Huygens-Fresnel})  and use    $e^{j\Phi_\text{irs}}=\exp(jk(\Phi_x x+\Phi_y y))$ and $\cos(\theta_{op})={z_p\over|\mathbf{r}_{op}|}$, which leads to
\iftoggle{Conf}{%
\begin{IEEEeqnarray}{rll}E_r(\mathbf{r}_{p})&= \frac{E_0z_p w_0}{j\lambda{w}(\tilde{d}_i)}\!\!\int\limits_{-\frac{L_y}{2}}^{\frac{L_y}{2}}  \!\!\int\limits_{-\frac{L_x}{2}}^{\frac{L_x}{2}} \!\!{\exp(jk|\mathbf{r}_{op}|)\over|\mathbf{r}_{op}|^2}  e^{jk(\Phi_x x+\Phi_y y)}\nonumber\\& \times e^{{-{\hat{x}^2+{\hat{y}}^2\over {w}^2(\tilde{d}_i)}} -j\left(k\left(\hat{d}_i+{\hat{x}^2+\hat{y}^2\over 2R(\tilde{d}_i)}\right)-\tan^{-1}\left(\frac{\tilde{d}_i}{z_0}\right)\right)}\mathrm{d}x\mathrm{d}y,\quad
\label{proofTheo1_1}
\end{IEEEeqnarray}
}{
\begin{IEEEeqnarray}{rll}E_r(\mathbf{r}_{p})&= \frac{E_0z_p w_0}{j\lambda{w}(\tilde{d}_i)}\!\!\int\limits_{-\frac{L_y}{2}}^{\frac{L_y}{2}}  \!\!\int\limits_{-\frac{L_x}{2}}^{\frac{L_x}{2}} \!\!{\exp(jk|\mathbf{r}_{op}|)\over|\mathbf{r}_{op}|^2}  e^{jk(\Phi_x x+\Phi_y y)} e^{{-{\hat{x}^2+{\hat{y}}^2\over {w}^2(\tilde{d}_i)}} -j\left(k\left(\hat{d}_i+{\hat{x}^2+\hat{y}^2\over 2R(\tilde{d}_i)}\right)-\tan^{-1}\left(\frac{\tilde{d}_i}{z_0}\right)\right)}\mathrm{d}x\mathrm{d}y,\quad
	\label{proofTheo1_1}
\end{IEEEeqnarray}
}
where we approximate $|\mathbf{r}_{op}|\approx d_p+x^2\left(\frac{1}{2d_p}+\frac{x_p^2}{2d_p^3}\right)+y^2\left(\frac{1}{2d_p}+\frac{y_p^2}{2d_p^3}\right)-{\frac{xx_p+yy_p}{d_p}}$ by assuming in  the exponential function $e^{jk|\mathbf{r}_{op}|}$, the following term is sufficiently small,
\begin{IEEEeqnarray}{rll}	
		k\frac{(x^2+y^2)^2}{8d_p^3}\ll 2\pi.
		\label{d_n}
	\end{IEEEeqnarray}
Substituting $x=x_e$ and $y=y_e$, we obtain
\begin{IEEEeqnarray}{rll}	
	\left(\frac{(x_e^2+y_e^2)^2}{8 \lambda}\right)^{1/3}\ll d_p,
	\label{d_n2}
\end{IEEEeqnarray}
and thus, for $d_n$ as defined in Theorem \ref{Theorem1}, our model is accurate for  $d_p\gg d_n$. For the other terms that include $|\mathbf{r}_{op}|$ in (\ref{proofTheo1_1}), we approximate  $|\mathbf{r}_{op}|\approx d_p$ and thus, we substitute  $\frac{z_p}{|\mathbf{r}_{op}|^2}=\frac{z_p}{d_p^2}$. Now,  we obtain from (\ref{proofTheo1_1}) that
\iftoggle{Conf}{%
	\begin{IEEEeqnarray}{rll}
		E_r(\mathbf{r}_{p})&=C_1  \int\limits_{-\frac{L_x}{2}}^{\frac{L_x}{2}} e^{-a{x^2}-bx} \mathrm{d}x \int\limits_{-\frac{L_y}{2}}^{\frac{L_y}{2}}  e^{-c{y^2}-dy} \mathrm{d}y,	
		\quad
		\label{seperatexy}
	\end{IEEEeqnarray}
}{
	\begin{IEEEeqnarray}{rll}
		E_r(\mathbf{r}_{p})&=C_1  \int\limits_{-\frac{L_x}{2}}^{\frac{L_x}{2}} e^{-a{x^2}-bx} \mathrm{d}x \int\limits_{-\frac{L_y}{2}}^{\frac{L_y}{2}}  e^{-c{y^2}-dy} \mathrm{d}y,	
		\label{seperatexy}
	\end{IEEEeqnarray}
}
where    $C_1=\frac{E_ow_0z_p}{j\lambda{w}(\tilde{d}_i)d_p^2}e^{jk(-d_i+d_p)-\sin^2(\theta_i)\nu x_0^2-\nu y_0^2+jkx_0\cos(\theta_i)+j\tan^{-1}\left(\frac{\tilde{d}_i}{z_0}\right)}$, $a=\sin^2(\theta_i)\nu-{jk}\left(\!\frac{1}{2d_p}+\frac{x_p^2}{2d_p^3}\!\right)$,
$b=-2x_0\sin^2(\theta_i)\nu+jk\left(\cos(\theta_i)+\frac{x_p}{d_p}-\Phi_x\right)$,
$c=\nu-{jk}\left(\frac{1}{2d_p}+\frac{y_p^2}{2d_p^3}\right)$,
$d=-2y_0\nu+jk\left(\frac{y_p}{d_p}-\Phi_y\right)$. Then,  using  \cite[Eq.~(2.33-1)]{integral} as follows
\begin{IEEEeqnarray}{rll}
	\int e^{-a{x^2}-bx} \mathrm{d}x=\frac{1}{2}\sqrt{\frac{\pi}{a}}\exp\left(\frac{b^2}{4a}\right)\text{erf}\left(\sqrt{a}x+\frac{b}{2\sqrt{a}}\right),\qquad
\end{IEEEeqnarray}
we can solve the integrals in (\ref{seperatexy}). Next, we transform  $\mathbf{r}_p$ from the $xyz$-coordinates to the lens plane coordinates.  Thus, ${\mathbf{r}}_r=	\mathbf{R}_{y_r}({\frac{\pi}{2}-\theta_r})\mathbf{R}_z({-\phi_r})\left(\mathbf{r}_p-\mathbf{r}_0\right)-\mathbf{r}_{r0}$, where $\mathbf{r}_{r0}=(0, 0, d_r)$ is the translation vector to the reflection plane and $\mathbf{R}_y(\cdot)$ and $\mathbf{R}_z(\cdot)$ are rotation matrices defined in the proof of Lemma \ref{Lemma1}.  Given the unitarity of matrices $\mathbf{R}_y(\cdot)$ and $\mathbf{R}_z(\cdot)$, ${\mathbf{r}_p}$ can be expressed as  ${\mathbf{r}_p}=\left(\mathbf{R}_{y_r}(\frac{\pi}{2}-{\theta_r})\mathbf{R}_z(-{\phi_r})\right)^T\left({\mathbf{r}_r}+\mathbf{r}_{r0}\right)+\mathbf{r}_0$, and thus,  any point ${\mathbf{r}_p}$ on the lens plane  is given by 
\iftoggle{Conf}{%
\begin{IEEEeqnarray}{rll}
		&\mathbf{r}_p
		=\mathbf{r}_0+\nonumber\\ &\!\!\begin{pmatrix}\!
			\cos(\phi_r)\sin(\theta_r) & \sin(\phi_r) & \cos(\phi_r)\cos(\theta_r)\\
			-\sin(\phi_r)\sin(\theta_r) & \cos(\phi_r) & -\sin(\phi_r)\cos(\theta_r)\\
			-\cos(\theta_r) & 0 & \sin(\theta_r)
		\end{pmatrix}
		\left(
		\mathbf{r}_r
		+
		\mathbf{r}_{r0}\right).
		\nonumber\\
		& \qquad
		\label{xyz}
\end{IEEEeqnarray}
}{
\begin{IEEEeqnarray}{rll}
	&\mathbf{r}_p
	=\mathbf{r}_0+\begin{pmatrix}
		\cos(\phi_r)\sin(\theta_r) & \sin(\phi_r) & \cos(\phi_r)\cos(\theta_r)\\
		-\sin(\phi_r)\sin(\theta_r) & \cos(\phi_r) & -\sin(\phi_r)\cos(\theta_r)\\
		-\cos(\theta_r) & 0 & \sin(\theta_r)
	\end{pmatrix}
	\left(
	\mathbf{r}_r
	+
	\mathbf{r}_{r0}\right).
	\label{xyz}
\end{IEEEeqnarray}
}
 Based on (\ref{xyz}) and since, $d_r\gg a, x_0, y_0$, we can approximate $d_p\approx d_r$, $z_p\approx d_r\sin(\theta_r)$, $\frac{x_p^2}{d_p^3}=\frac{1}{d_r}\cos^2(\phi_r)\cos^2(\theta_r)$, and $\frac{y_p^2}{d_p^3}=-\frac{1}{d_r}\sin^2(\phi_r)\cos^2(\theta_r)$.  Finally, applying  transformation  (\ref{xyz})  in (\ref{seperatexy}) and (\ref{d_n2}), we obtain (\ref{theo1}).  This completes the proof.

 \section{Proof of Theorem~\ref{Theorem4}}\label{App4}
First, we substitute (\ref{theo1}) into (\ref{irradiance}) and since given the size of the lens, $\frac{x_r}{d_r}, \frac{y_r}{d_r}\ll 1$, we approximate $x_r=y_r\approx a_0$ in the  $\text{erf}(\cdot)$  terms in  (\ref{theo1})  and substitute  by  $C_2$. After some simplification, we obtain
 \begin{IEEEeqnarray}{rll}
 	&h_{\text{irs}}=\frac{\pi^2{|C_2|^2}E_0^2 {w_0^2}\sin^2(\theta_r)}{32\eta{|b_x|}{|b_y|}{\lambda^2{w}^2(\tilde{d}_i)d_r^2}} e^{-2\nu\left(\sin^{\!2}\!(\theta_i) x^2_{0}+ y^2_{0}\right)}  \times\nonumber\\
 	&\!\!\int\limits_{-a_0}^{a_0}\!\int\limits_{-\epsilon}^{\epsilon}   \!\exp\left(\!\!\frac{-k^2}{2\tilde{b}_x}\!\!\left(\bar{X}-\Phi_x\right)^2\!\!\right)\!\exp\left(\!\!\frac{-k^2}{2\tilde{b}_y}\!\!\left(\bar{Y}-\Phi_y\right)^2\!\!\right)\mathrm{d}y_r \,\mathrm{d}x_r,\!\qquad
 	\label{ptheo42}
  \end{IEEEeqnarray}
 where  $\bar{X}=\varphi_x+c_1x_r+c_2y_r$,  $\bar{Y}=\varphi_y+c_3x_r+c_4y_r$, and $\epsilon=\sqrt{a_0^2-x_r^2}$. Next, given the small size of the circular lens, we  approximate it  by  a  square with the same area and length $\sqrt{\pi}a_0$ and rewrite (\ref{ptheo42}) as follows
  \begin{IEEEeqnarray}{rll}
 	h_{\text{irs}}&={C}_h\int\limits_{-\varepsilon}^{\varepsilon}\int\limits_{-\varepsilon}^{\varepsilon}   e^{-\left(\rho_xx_r^2+\rho_yy_r^2+\rho_{xy}x_r y_r+\varrho_xx_r+\varrho_yy_r\right)}\mathrm{d}x_r \, \mathrm{d}y_r.\!\qquad
 	\label{ptheo43}
 \end{IEEEeqnarray}
Then, we solve the inner integral by applying \cite[Eq.~(2.33-1)]{integral}, which leads to (\ref{theo4}) for out-of-plane reflection and  completes the proof.

\bibliographystyle{IEEEtran}
\bibliography{My_Citation_1-07-2020}
\end{document}